\documentclass[twocolumn,showpacs,preprintnumbers,amsmath,amssymb,pra]{revtex4}

\usepackage[dvips]{graphicx}
\usepackage{dcolumn}
\usepackage{bm}

\newcommand{\imagi} {\mathrm{i}}
\newcommand{\euler} {\mathrm{e}}
\newcommand{\vh} {v_{\mathrm{h}}}
\newcommand{\vx} {v_{\mathrm{x}} }
\newcommand{\vc} {v_{\mathrm{c}}}
\newcommand{\vhx} {v_{\mathrm{hx}}}
\newcommand{\gxc} {g_{\mathrm{xc}}}
\newcommand{\del} {\mathrm{d}}
\newcommand{\Hop} {\widehat{H}}
\newcommand{\Psifull} {\psi ( x_{1}, x_{2}, t  )}
\newcommand{\Psifullend}  {\psi ( x_{1}, x_{2}, T  )}
\newcommand{\Psifullsq}{\vert \Psifull \vert^{2}}
\newcommand{\Psitwoplus}{\psi^{(2 \protect \raisebox{0.25ex}{$\scriptscriptstyle +$})}}
\newcommand{\phiplus}{\phi^{(\protect\raisebox{0.25ex}{$\scriptscriptstyle +$})}}
\newcommand{\rhotwoplus}{\rho^{(2 \protect\raisebox{0.25ex}{$\scriptscriptstyle +$})}}
\newcommand{\ntwoplus}{n^{(2 \protect\raisebox{0.25ex}{$\scriptscriptstyle +$})}}
\newcommand{\ofx}[1][]{( x_{#1} )}
\newcommand{\ofxt}[1][]{( x_{#1},t )}
\newcommand{\ofkt}[1][]{( k_{#1},t )}
\newcommand{\oft}{( t )}
\newcommand{\ofT}{( T )}
\newcommand{\ofxx}{( x_{1}, x_{2} )}
\newcommand{\ofxxt}{( x_{1}, x_{2}, t )}
\newcommand{\ofkkt}{( k_{1}, k_{2}, t )}
\newcommand{\ofkkT}{( k_{1}, k_{2}, T )}
\newcommand{\Ptwoplus}{P^{2 \protect\raisebox{0.25ex}{$\scriptscriptstyle +$}}}
\newcommand{\Heplus}{\mathrm{He}^{\protect\raisebox{0.25ex}{$\scriptscriptstyle +$}}}
\newcommand{\Hetwoplus}{\mathrm{He}^{2 \protect\raisebox{0.25ex}{$\scriptscriptstyle +$}}}
\newcommand{\nm}{\mathrm{nm}}
\newcommand{\LK}{\mathrm{\scriptscriptstyle LK05}}
\newcommand{\Ion}{\mathrm{\scriptscriptstyle Ion}}

\newcommand{\PP}{\mathrm{\scriptscriptstyle PP}}
\newcommand{\KS}{\mathrm{\scriptscriptstyle KS}}
\newcommand{\epsen}{\epsilon_{\mathrm{en}}}
\newcommand{\epsee}{\epsilon_{\mathrm{ee}}}
\newcommand{\Ip}[1][]{I_{\mathrm{p}}^{\mbox{$\scriptscriptstyle (#1)$}}}

\begin{document}
\preprint{APS}

\title{Momentum distributions in time-dependent density functional theory:\\ Product phase approximation for non-sequential double ionization\\ in strong laser fields}

\author{F.~Wilken}
\author{D.~Bauer}
\affiliation{Max-Planck-Institut f\"ur Kernphysik, Postfach 103980,
69029 Heidelberg, Germany}

\date{\today}

\begin{abstract}
We investigate the possibility to deduce momentum space properties from time-dependent density functional calculations. Electron and ion momentum distributions after double ionization of a model Helium atom in a strong few-cycle laser pulse are studied. We show that, in this case, the choice of suitable functionals for the observables is considerably more important than the choice of the correlation potential in the time-dependent Kohn-Sham equations. By comparison with the solution of the time-dependent Schr\"odinger equation, the insufficiency of functionals neglecting electron correlation is demonstrated. We construct a functional of the Kohn-Sham orbitals, which in principle yields the exact momentum distributions of the electrons and the ion. The product-phase approximation is introduced, which reduces the problem of approximating this functional significantly.
\end{abstract}

\maketitle

\section{Introduction}
\label{Section Introduction}
Time-dependent density functional theory (TDDFT)  \cite{RungeGross1984} is a remarkably successful approach to the study of many-body systems in time-dependent external fields \cite{Marques2006}. The essential statement of TDDFT is the same as that of the well-established ground state density functional theory (DFT) \cite{HohenbergKohn1964}: all observables are, in principle, functionals of the particle density alone. Since the latter is always a three-dimensional entity, independent of the number of particles involved, the computational cost of actual (TD)DFT calculations scales exponentially more favorable than the solution of the many-body (time-dependent) Schr\"odinger equation.

In practice, almost all (TD)DFT calculations are performed using the (time-dependent) Kohn-Sham scheme [(TD)KS] (see, e.g., \cite{Marques2006}) where the density is calculated with the help of auxiliary, non-interacting particles moving in an effective potential. The ``art'' of (TD)DFT is two-fold, namely finding sufficiently accurate approximations to the density functionals of (i) the unknown effective potential and (ii) the observables of interest. Fortunately, for many practical applications both items are uncritical \cite{Marques2006}. An example is the calculation of the optical response of bio-molecules where even the simple local density approximation of the effective potential yields reasonable results, and the observable can be calculated from a known and explicit functional of the density (the time-dependent dipole).  

However, when it comes to the correlated motion of a {\em few} particles in a strongly driven system, TDDFT faces major challenges. In that respect, non-sequential double ionization (NSDI) serves as the ``worst case'' scenario for TDDFT. Theoretically, NSDI was addressed successfully using the strong-field approximation (see, e.g., \cite{BeckerDoerner2005} and references therein) and classical methods \cite{FuLiu2001,HoPanfili2005}. The widely accepted mechanism behind NSDI relies on the rescattering of the first electron with its parent ion, collisionally ionizing (or exciting) the second electron. 

In the recent publications Refs.~\cite{LeinKuemmel2005,WilkenBauer2006} significant progress was made in the treatment of NSDI within TDDFT as far as ionization {\em yields} are concerned. The latter display as a manifestation of the electron-electron correlation involved in NSDI the celebrated ``knee'' structure in the double ionization yield, which was, until recently, not being reproduced within TDDFT. Reference \cite{LeinKuemmel2005} addressed issue (i) above (the effective potential) while Ref.~\cite{WilkenBauer2006} focused on item (ii), the functional for the observable ``double ionization''. It was shown that (i) taking the derivative discontinuities at integer bound electron numbers into account and (ii) using an adiabatic approximation for the correlation function needed to calculate the double ionization probability, the NSDI ``knee'' can be reproduced.

In our current work we turn to the much harder problem of momentum distributions (or energy spectra \cite{VeniardTaieb2003}). In the NSDI regime the ion momentum spectra, as measured in experiments employing ``reaction microscopes'' (see, e.g., \cite{UllrichMoshammer2003,BeckerDoerner2005}), show a characteristic ``double-hump'' structure, i.e., maxima at non-vanishing ion momenta. The maxima at non-zero ion momenta are easy to understand within the rescattering scenario mentioned above: the first electron preferentially returns to the ion, collisionally ionizing the second electron, at times when the vector potential of the laser field is non-zero. Since the vector potential at the ionization time equals the final drift momentum at the detector, non-vanishing electron momenta (and, due to momentum conservation, non-vanishing ion momenta) are likely.  In a TDKS treatment of NSDI in He starting from a spin-singlet state, the rescattering scenario is ``hidden'' in a single, spatial Kohn-Sham (KS) orbital. As we shall demonstrate, taking the auxiliary KS particles for real electrons and Fourier-transforming their position space product wavefunction to momentum space leads to ion momentum spectra in very poor agreement with the exact ones. A better approximation to calculate correlated electron momentum spectra in the NSDI regime is required. With the present paper we aim at contributing to this goal by showing that item (ii) above, namely the construction of the functional for the observable, is the critical issue while (i) known effective potentials are sufficient, at least at the current level of accuracy.

In Sec.\,\ref{Section Model System} the model Helium system used to study the ionization process and the ensuing momentum distributions is introduced. In Sec.\,\ref{Section Momentum Densities} the method to calculate electron and ion momentum distributions is explained. Results from the solution of the time-dependent Schr\"odinger equation (TDSE) in Sec.\,\ref{Section Momentum Distribution from TDSE} serve as a reference for the results obtained using TDDFT in Sec.\,\ref{Section Momentum Distributions from TDDFT}: The insufficiency of uncorrelated functionals to calculate electron and ion momentum distributions (Sec.\,\ref{Section Momentum Uncorrelated Functionals}) and the relative insignificance of the correlation potential (Sec.\,\ref{Section The Role of the Correlation Potential}) lead us to the construction of correlated functionals in Sec.\,\ref{Section Momentum Towards a Correlated Functional}. In Sec.\,\ref{Section Momentum Product Phase Approximation} we introduce the product-phase approximation, which reduces the problem of approximating the correlated functionals to that of approximating the exchange-correlation function.

For consistency, we restrict ourselves to the presentation of results for laser pulses with $\lambda \! = \! 780\,\nm$ and $N \! = \!3$ cycles. We stress, however, that the general conclusions drawn hold also for $\lambda \! = \! 614 \,\nm$, $N \! = \!3$ and $\lambda \! = \! 780 \,\nm$, $N \! = \!4$ laser pulses, as we have checked explicitly.

\section{Model System}
\label{Section Model System}

A Helium atom exposed to linearly polarized laser pulses with $N=3$ cycles and $\sin^{2}$-pulse envelopes is investigated. The length of the pulses with a frequency of $\omega=0.058$ (corresponding to the experimentally used $\lambda=780 \, \nm$) is $T \! = \! 2\, N \pi / \omega$, and the vector potential reads $A \oft = \hat{A} \sin^{2} \left( \frac {\omega}{2 N} \, t \right) \, \sin ( \omega \, t )$ for $0 \leq t \leq T$ and zero otherwise (atomic units are used unless otherwise indicated). We use the dipole approximation, i.e., the spatial dependence of the laser field is neglected. The linear polarization of the laser pulse thus allows us to describe the system by a one-dimensional model Helium atom with soft-core potentials for the Coulomb interactions. It is known that the essential features of the ionization process are described well by this model \cite{Bauer1997,LappasLeeuwen1998,LeinGross2000,DahlenLeeuwen2001,LeinKuemmel2005,WilkenBauer2006}. Initially, the electrons are assumed to occupy the spin-singlet groundstate of Helium, and due to the neglect of magnetic effects in the dipole approximation the electrons stay in the spin-singlet state during the interaction with the laser pulse. Thus it is sufficient to study the spatial wavefunction, which has to be symmetric under exchange of the electrons. 

The TDSE $\imagi \partial_{t} \, \psi \ofxxt = \Hop \ofxxt \, \psi \ofxxt$ is solved for laser pulses with different effective peak intensities $I=I ( \hat A )$. A trivial gauge-transformation cancels the purely time-dependent $A^{2}$-term and yields the Hamiltonian
\begin{equation}
\Hop = \sum_{i=1,2} \Big( - \frac{1}{2} \, \partial^{2}_{x_{i}} + V ( x_{i}, t ) \Big) + W ( \vert x_{1}-x_{2} \vert ) \, ,
\label{Formula TDSE Hamiltonian}
\end{equation}
with $\Hop = \Hop \ofxxt$. The external potential is $V \ofxt = - \imagi \, A ( t ) \, \partial_{x} - 2 / \sqrt{ x^{2}+\epsen}$, the electron-electron interaction potential is given by $W \ofx = 1 / \sqrt{ x^{2} +\epsee}$. The soft-core parameters $\epsen$ and $\epsee$ are chosen to yield the correct ionization potentials. Reproducing the ionization potential of $\Heplus$, $\Ip[2]=2.0$ in a corresponding model $\Heplus$ ion fixes $\epsen=0.5$. The choice $\epsee=0.329$ yields the ionization potential of Helium, $\Ip[1] = 0.904$. All results presented in this work are qualitatively insensitive to the precise values of the soft-core parameters. 

As the two electrons constitute a spin-singlet state for all times they are described by the same KS orbital. Therefore, in a TDDFT treatment, we have only one time-dependent Kohn-Sham equation (TDKSE) $ \imagi \, \partial_{t} \, \phi \ofxt = \Hop^{\KS} \ofxt \, \phi \ofxt$ with the Hamiltonian
\begin{equation}
\Hop^{\KS} \ofxt = - \frac{1}{2} \, \partial^{2}_{x} + V \ofxt+ \vhx \ofxt + \vc \ofxt \, .
\label{Formula Model Hamiltonian KS}
\end{equation}
The Hartree-exchange potential  $\vhx =\vh +\vx$ follows as $\vhx \ofxt = \frac{1}{2} \int \del x' \, n ( x',t ) / \sqrt{ ( x- x' )^{2} + \epsee^{\KS} }$. We have used the exact exchange term for Helium $\vx \ofxt = - \vh \ofxt / \, 2$, which is local as both electrons are described by the same orbital. 

Setting $\vc=0$ yields, in the special case of the Helium atom or He-like ions, an identical description as the time-dependent Hartree-Fock (TDHF) treatment (due to the locality of $\vx$). The LK05 potential $\vc^{\LK}$ \cite{LeinKuemmel2005} takes into account the discontinuous change in the correlation potential when the number of bound electrons $N \oft = \int_{-a}^{+a} \del x \, n \ofxt$ passes integer numbers, $\vc^{\LK} \ofxt =  \left[ B \oft / \left( 1+\mathrm{exp} [ C (B \oft -2) ] \right) -1 \right] \vhx \ofxt $, where $C$ is a sufficiently large constant (we set $C=50$) and $B \oft =N_{0}/N \oft$. In order to encompass all bound states the parameter $a$ is chosen as $a=6 \, \mathrm{a.u.}$ throughout this work, results being insensitive to the precise value of $a$. We use $\epsee^{\KS}=0.343$ in the Hartree-exchange potential $\vhx$ to acquire $\Ip[1]=0.904$ for the model Helium atom. The TDSE and TDKSE are solved by a split-operator time propagator on a numerical grid (see, e.g., \cite{BauerKoval2006} and references therein).

Along the lines of Ref.\,\cite{LeinKuemmel2005} we construct from the TDSE solution an exact KS orbital (EKSO). The Schr\"odinger solution gives the exact density of our model Helium atom
$ n \ofxt  = 2\, \int \del x_{2}  \, \vert \psi ( x, x_{2},t ) \vert^{2} = 2\, \int \del x_{1}  \, \vert \psi  ( x_{1}, x,t ) \vert^{2}$ and the exact probability current $j \ofxt$. From the equality of the exact and KS currents in the case of a one-dimensional system, the phase of the EKSO is determined as $\vartheta \ofxt  =  \int_{-\infty}^{x} \del x' \, j (x',t) / n (x',t) + \alpha \oft$. The unknown purely time-dependent phase factor $\alpha \oft$ does not affect the results presented in this work and is therefore set to zero. The EKSO $\phi \ofxt = \sqrt{ n\ofxt / \, 2} \, \euler^{\imagi \, \vartheta \ofxt}$ is thus identical to the orbital a TDDFT calculation with the exact correlation potential $\vc$ would yield via the TDKS scheme. The EKSO allows us to separate the challenges facing TDDFT calculations (cf.\ Sec.\,\ref{Section Introduction}):  finding (i) a suitable approximation of $\vc$ (where it serves as a reference for the resulting orbital) and (ii) appropriate functionals for observables (where it is the exact input).

\section{Momentum Densities}
\label{Section Momentum Densities}

We partition the two-electron space and associate with single ionization the area $\mathcal{A} \, ( \Heplus ) = \{ ( x_{1},x_{2} ) \mid \vert x_{i} \vert > a, \vert x_{j \neq i} \vert \leq a \ \forall \ i,j \in \{1,2\} \}$ and with double ionization the area $\mathcal{A} \, ( \Hetwoplus ) = \{ ( x_{1},x_{2} ) \mid \vert x_{1} \vert > a, \vert x_{2} \vert > a \}$. Integrating $\Psifullsq$ over these areas then yields the respective ionization probabilities, with the double ionization probability given by
$\Ptwoplus \oft = \int \! \! \! \int_{\mathcal{A} \, (\Hetwoplus)} \del x_{1} \, \del x_{2} \, \Psifullsq$. This scheme to determine ionization probabilities from the two-electron wavefunction has been successfully used in numerous similar calculations \cite{Bauer1997,LappasLeeuwen1998,DahlenLeeuwen2001}.

The wavefunction $\psi \ofxxt$ can be described equivalently in momentum space by its Fourier transform
$\left( 2 \, \pi \right) \, \psi \ofkkt = \int \del x_{1} \int \del x_{2} \, \Psifull \, \euler^{- \imagi \, ( k_{1} \, x_{1} + k_{2} \, x_{2} )}$. As the wavefunction in momentum space is normalized to one, the pair density in momentum space is given by $\rho \ofkkt = 2 \, \vert \psi \ofkkt \vert^2$. 

At times $0 < t < T$ during the laser pulse the velocity of the electrons is actually given by $ \dot{x}_{i} \oft = k_{i} \oft +A \oft$, i.e., the sum of the canonical momentum $k_{i}$ and the value of the vector potential at the respective time. In this work we investigate properties of the system at $t=T$ after the laser pulse. As $A \ofT \! = \! 0$, canonical momenta $k$ and drift momenta are identical.

We are interested mainly in the double ionization process and thus Fourier transform only the wavefunction in the area $\mathcal{A} \, ( \Hetwoplus )$ associated with double ionization. The resulting sharp step at the boundary of $\mathcal{A} \, ( \Hetwoplus )$ at $\vert x_{i} \vert = a, \ \vert x_{j \neq i} \vert \geq a$, with $i,j \in \{ 1,2 \}$, is a potential source of artifacts when Fourier transformed. Hence, a smoothing function $f \ofxx = \prod_{i=1}^{2} 1 / \sqrt{ 1+\euler^{-c\, \vert x_{i}-a \vert} }$  is introduced. The factor $c$ has to be of the order of one, in this work we choose $c \! = \! 1.25$. The smoothing function is constructed so that $\int \! \! \int \del x_{1} \del x_{2} \, f^{2} \ofxx \, b = \int \! \! \int_{\mathcal{A} \, (\Hetwoplus)} \del x_{1} \, \del x_{2} \, b$ for a constant $b$. This condition ensures that the wavefunction $\Psitwoplus \ofxxt = f \ofxx \, \psi \ofxxt$ gives to a good approximation the same double ionization probability as the original wavefunction, i.e., that $\int \del x_{1} \int \del x_{2} \, f^{2} \ofxx \vert \psi \ofxxt \vert^2 \simeq \Ptwoplus$. 
The correlated wavefunction of the electrons freed in double ionization in momentum space is thus calculated as
\begin{eqnarray}
\lefteqn{ \left( 2 \, \pi \right) \, \Psitwoplus \ofkkt = } & & \nonumber \\
 &  & \int \del x_{1} \! \int \del x_{2} \, \psi^{(2 \protect\raisebox{0.25ex}{$\scriptscriptstyle +$})} \ofxxt \, \euler^{- \imagi \, ( k_{1} \, x_{1} + k_{2} \, x_{2} )} \, .
\label{Formula Momentum Psitwoplus}
\end{eqnarray}
This approach is equivalent to projecting out the states corresponding to single and no ionization and is known to lead to accurate momentum distributions \cite{LeinGross2000}.

From the wavefunction we construct the momentum pair density of the electrons freed in double ionization
\begin{equation}
\rhotwoplus \ofkkt = 2 \, \vert \Psitwoplus \ofkkt \vert ^2 \, .
\label{Formula Momentum Pair Density}
\end{equation}
The probability to find at time $t$ an electron freed in double ionization with momentum $k_{1}$ in $\del k_{1}$ and an electron with $k_{2}$ in $\del k_{2}$ is then $\rhotwoplus \ofkkt \, \del k_{1} \del k_{2}$.

In experiments, it is easier to measure the momentum of the $\Hetwoplus$ ion $k_{\Ion}$ after double ionization instead of individual electron momenta. As the total photon momentum involved is negligibly small, this provides information about the sum of the electron momenta via momentum conservation 
$k_{1}+k_{2}=-k_{\Ion}$. The ion momentum density then follows from the momentum pair density of the electrons freed in double ionization (\ref{Formula Momentum Pair Density}) as
\begin{eqnarray}
\ntwoplus_{\Ion} ( k_{\Ion},t ) & = & \frac{1}{2} \, \int \del k \, \rhotwoplus ( -k_{\Ion} \! -k, k, t ) \nonumber \\
& = & \frac{1}{2} \, \int \del k \, \rhotwoplus ( k, -k_{\Ion} \! -k, t ) \, ,
\label{Formula Momentum Ion Density}
\end{eqnarray}
due to the symmetry of the electron momentum pair density. The factor $1/2$ ensures the correct normalization since the system consists of only one ion but two electrons.
The ion momentum density $\ntwoplus_{\Ion} ( k_{\Ion},t ) \, \del k_{\Ion}$ gives the probability to find at time $t$ the $\Hetwoplus$ ion with momentum $k_{\Ion}$ in $\del k_{\Ion}$.

\section{Momentum Distributions from the TDSE}\label{Section Momentum Distribution from TDSE}
From the numerical solution of the TDSE we obtain $\Psifullend$ after the interaction with the laser pulse. In the left hand side of Fig.\,\ref{Figure Momentum Pair Density TDSE-EKSO} the momentum pair density of the electrons freed in double ionization, as calculated from Eq.\,(\ref{Formula Momentum Pair Density}), is shown. 

For all but the highest intensity depicted, electrons have the highest probability to move at different velocities $\vert k_{1} \vert \neq \vert k_{2} \vert$ ($ \dot{x}_{i} \ofT =k_{i} \ofT$ since $A \ofT \! = \! 0$, cf.\ discussion in Sec.\,\ref{Section Momentum Densities}) but in the same direction ($\mathrm{sgn}(k_{1})=\mathrm{sgn}(k_{2})$). Depending on the laser intensity the probability for the double ionization process is highest at different half-cycles of the laser pulse, i.e., different signs of the vector potential. Therefore, the favored direction in which the electrons leave the atom varies with intensity. NSDI can be understood by a recollision mechanism where one electron returns to the $\Heplus$ ion and frees the second electron (see, e.g., \cite{BeckerDoerner2005}). The results of the TDSE then imply that both electrons leave the atom in the same direction but due to Coulomb repulsion their velocities differ, in accordance with earlier results for a longer laser pulse \cite{LeinGross2000}.

The ``butterfly'' shape of the momentum pair density of the electrons freed in double ionization as shown in Fig.\,\ref{Figure Momentum Pair Density TDSE-EKSO} is evidence that it is highly correlated, as it cannot be reproduced by multiplying two orbitals for the respective electrons. 

For $I \! = \! 6.96 \times 10^{15}\, \mathrm{W/cm^{2}}$ both electrons have the highest probability to leave the atom in the same direction with similar velocities $k_{1} \approx k_{2}$. This can only be the case when the Coulomb repulsion between the electrons is weak, i.e., when they are removed sequentially, resulting in a large spatial separation. The final non-vanishing velocities are due to the high intensity of the laser pulse, which ionizes the atom so rapidly that $A \oft \neq 0$ when the first electron is freed. The grid-like structure typical for a product wavefunction is seen, the electron correlation being weak.

From the momentum pair density  of the electrons freed in double ionization $\rhotwoplus \ofkkT$ (\ref{Formula Momentum Pair Density}) we calculate the ion momentum density $\ntwoplus_{\Ion} \ofkkT$  (\ref{Formula Momentum Ion Density}). For different effective peak intensities the density of the ion momentum is depicted in Fig.\,\ref{Figure Ion Momentum Density TDSE-EKSO}. It exhibits peaks at non-zero momenta. As explained in Sec.\,\ref{Section Introduction}, these are typical for recollision processes when the first freed electron recollides close to the maximum of the vector potential, i.e., when $\vert A \oft \vert \approx \widehat A$. Hence, the sum of the momenta of both electrons is non-zero, and, by momentum-conservation, this holds for the ion momentum as well \cite{RudenkoZrost2004}.

For an infinitely long laser pulse of laser period $T/N$, $\Hop (t+T/N) = \Hop (t)$ holds while this symmetry is broken in the case of few-cycle laser pulses. Hence, with respect to the dislodged electrons there is no spatial inversion symmetry, leading to asymmetric ion momentum distributions \cite{LiuRottke2004,RottkeLiu2006,FariaLiu2004}. This effect is clearly seen in Fig.\,\ref{Figure Ion Momentum Density TDSE-EKSO}. For the three lowest intensities a process with $k_{\Ion} \geq 0$ dominates while with increasing intensities processes with $k_{\Ion} \leq 0$ become more likely. In addition, a central peak gets more and more pronounced, showing that the relative probability of sequential double ionization increases. The fact that the peak is not centered around $k_{\Ion}=0$ for $I \! = \! 6.96 \times 10^{15}\, \mathrm{W/cm^{2}}$ is again due to the high intensity and the short duration of the laser pulse, as explained above.

\section{Momentum Distributions from TDDFT}
\label{Section Momentum Distributions from TDDFT}
DFT can be formulated in momentum space (see, e.g., \cite{DreizlerGross1999}), and this seems to be the obvious path to follow when one is interested in the calculation of momentum spectra. However, momentum space DFT lacks the ``universality'' feature of the Hohenberg-Kohn theorem \cite{HohenbergKohn1964}, meaning that each system under study requires a different momentum space effective potential---an entirely unattractive feature. We therefore prefer to make the ``detour'' via standard, universal, position space TDDFT. In the case of single ionization, a straightforward calculation of the momentum or energy spectrum from the Fourier-transformed valence KS orbital may be a good approximation (see, e.g., the approach followed in Ref.~\cite{PohlReinhard2000}).  Instead, it is less obvious how to determine {\em correlated} momentum spectra from position space TDKS orbitals.  
 
As explained in the Introduction, determining momentum pair densities and ion momentum densities from a TDDFT approach faces two challenges: The first is to find an approximate correlation-potential $\vc$ in the TDKSE to reproduce the exact density $n \ofxt$ with sufficient accuracy. The second, more difficult one, amounts to assign a suitable functional of the density to the respective observable. As both the ion momentum density and the momentum pair density (via their probability interpretations, cf.\ Sec.\,\ref{Section Momentum Densities}) are observables, the Runge-Gross theorem assures that functionals of the density alone exist \cite{RungeGross1984}.


\subsection{Uncorrelated functionals}
\label{Section Momentum Uncorrelated Functionals}
Treating the KS orbital as if it were a one-electron wavefunction yields a product wavefunction $\phi \ofxt[1] \, \phi \ofxt[2]$. This is the same assumption frequently made to derive uncorrelated ionization probability functionals (see Ref.\,\cite{WilkenBauer2006} and references therein). 

The Fourier transformed KS orbital for $\vert x \vert > a$, i.e., with the bound states projected out (see Sec.\,\ref{Section Momentum Densities}) is
\begin{equation}
\sqrt{ 2 \, \pi } \, \phiplus \ofkt = \int \del x \, f \ofx \, \phi \ofxt \, \euler^{- \imagi \, k \, x} \, ,
\label{Formula Momentum Phiplus}
\end{equation}
with $f \ofx = 1 / \sqrt{ 1+\euler^{-c\, \vert x-a \vert}}$ the one-dimensional smoothing function equivalent to the smoothing function used in Sec.\,\ref{Section Momentum Densities}. 
\begin{figure}[!tb]
\centering
\includegraphics[angle=0, width=0.45\textwidth]{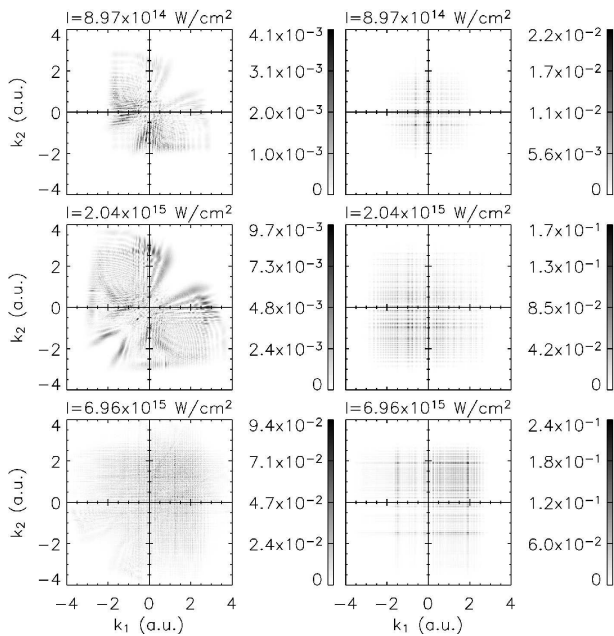}
\caption{ Contour plots of the momentum pair density $\rho^{2 {\scriptscriptstyle +}} \ofkkT$ of the electrons freed in double ionization. Results calculated from the uncorrelated functional (\ref{Formula Momentum Pair Density Uncorrelated}) using the EKSO (right hand side) are compared to the TDSE (left hand side) solution. Momentum pair densities for $\lambda \! = \! 780\,\nm$, $N \! = \!3$-cycle laser pulses with different effective peak intensities are shown.\label{Figure Momentum Pair Density TDSE-EKSO}}
\end{figure}
Calculating the momentum pair density (\ref{Formula Momentum Pair Density}) and the ion momentum density (\ref{Formula Momentum Ion Density}) from the product wavefunction gives the uncorrelated functional for the momentum pair density of the electrons freed in double ionization
\begin{equation}
\rhotwoplus \ofkkt = 2 \, \vert \, \phiplus \ofkt[1] \, \phiplus \ofkt[2] \vert^{2}
\label{Formula Momentum Pair Density Uncorrelated}
\end{equation}
and the uncorrelated functional for the ion momentum density of $\Hetwoplus$ 
\begin{equation}
\ntwoplus_{\Ion} ( k_{\Ion},t ) = \int \del k \, \vert \, \phiplus ( -k_{\Ion} \! -k, t ) \, \phiplus \ofkt \vert ^{2} \, .
\label{Formula Momentum Ion Momentum Density Uncorrelated}
\end{equation}
Equations (\ref{Formula Momentum Pair Density Uncorrelated}) and (\ref{Formula Momentum Ion Momentum Density Uncorrelated}) are not functionals of the density alone but due to the Fourier transformation they are dependent on the density and on the phase of the KS orbital.

The momentum pair density at $t=T$, as calculated from the uncorrelated functional (\ref{Formula Momentum Pair Density Uncorrelated}) using the EKSO, is depicted in the right part of Fig.\,\ref{Figure Momentum Pair Density TDSE-EKSO} for $\lambda \! = \! 780\,\nm$, $N \! = \!3$-cycle laser pulses with different intensities. Comparison with the left hand side showing the momentum pair density calculated from the correlated Schr\"odinger wavefunction $\psi (x_{1},x_{2},T)$ confirms that only for the highest intensity a product wavefunction approach is reasonable. For lower intensities the uncorrelated functional for the momentum pair density does not exhibit the typical ``butterfly''-shaped correlation structures of the Schr\"odinger solution. Instead, the grid-like structure typical for a product wavefunction is clearly visible.

\begin{figure}[!tb]
\centering
\includegraphics[angle=0, width=0.45\textwidth]{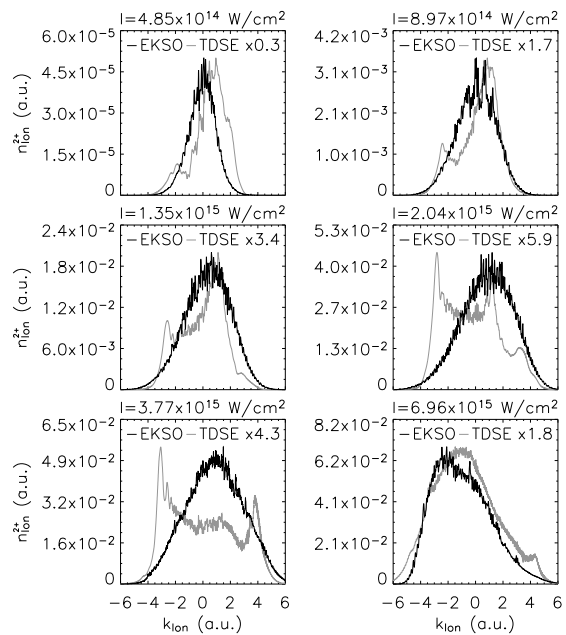}
\caption{ Ion momentum density of the model $\Hetwoplus$ ion after interaction with $\lambda \! = \! 780\,\nm$, $N \! = \!3$-cycle laser pulses with different effective peak intensities. The density calculated using the EKSO in the uncorrelated functional (\ref{Formula Momentum Ion Momentum Density Uncorrelated}) is compared to results from the TDSE. \label{Figure Ion Momentum Density TDSE-EKSO}}
\end{figure}

For the same system we calculate from Eq.\,(\ref{Formula Momentum Ion Momentum Density Uncorrelated}) the ion momentum density using the EKSO. In Fig.\,\ref{Figure Ion Momentum Density TDSE-EKSO} the $\Hetwoplus$ ion momentum density is compared to the results from the TDSE, which are scaled to enable the comparison of qualitative features. The different values of the integrals over the ion momentum densities are due to the different double ionization probabilities, as can be seen from $\int \del k_{\Ion} \, \ntwoplus_{\Ion} \ofkt[\Ion] \simeq \Ptwoplus$, which follows from Eq.\,(\ref{Formula Momentum Psitwoplus}) (see Ref.\,\cite{WilkenBauer2006} and references therein for a discussion of this particular problem). Apart from the highest intensity the density is centered around a central peak at $k_{\Ion} \approx 0$. This is evidence that correlations, which are not included in the uncorrelated functionals for the observables, are responsible for the distinct peaks of the ion momentum density at non-zero momenta. This result is consistent with the analysis of the results of the TDSE (Sec.\,\ref{Section Momentum Distribution from TDSE}), which attributes the peaks at $k_{\Ion} \neq 0$ to electron rescattering, i.e., to an interaction between the electrons. For the highest intensity shown in Fig.\,\ref{Figure Ion Momentum Density TDSE-EKSO}, sequential double ionization becomes dominant (cf.\ Sec.\,\ref{Section Momentum Distribution from TDSE}), so that the description using the EKSO in the uncorrelated functional reproduces the ion momentum density reasonably well.

\subsection{The role of the correlation potential}
\label{Section The Role of the Correlation Potential}
To underline the importance of the functional for the ion momentum density we use the correlation potentials $\vc=0$ (TDHF) and $\vc^{\LK}$(LK05) in the TDKSE for our model He atom interacting with the $\lambda \! = \! 780\,\nm$, $N \! = \!3$-cycle laser pulses (cf.\ Sec.\,\ref{Section Model System}). 

\begin{figure}[!tb]
\centering
\includegraphics[angle=0, width=0.45\textwidth]{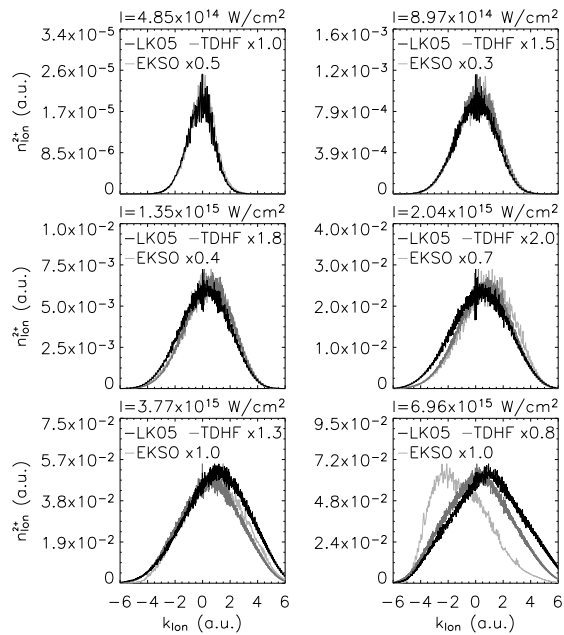}
\caption{ Ion momentum density of the model $\Hetwoplus$ ion after interaction with $\lambda \! = \! 780\,\nm$, $N \! = \!3$-cycle laser pulses with different effective peak intensities. The densities are calculated from the uncorrelated functional (\ref{Formula Momentum Ion Momentum Density Uncorrelated}) using the EKSO and the orbitals obtained with $\vc=0$ (TDHF) and $\vc^{\LK}$ (LK05).\label{Figure Ion Momentum Density EKSO-TDHF-LK}}
\end{figure}

In Fig.\,\ref{Figure Ion Momentum Density EKSO-TDHF-LK} the ion momentum densities obtained from using the respective orbitals in the uncorrelated functional for the ion momentum density (\ref{Formula Momentum Ion Momentum Density Uncorrelated}) are compared to the results with the EKSO, i.e., the orbital which the exact $\vc$ would yield. For the TDHF approach, results are similar to the results using the LK05-potential. Both approximations lead to uncorrelated ion momentum densities  which are close in qualitative terms to the EKSO results. Only at the highest intensity $I = 6.96 \times 10^{15} \, \mathrm{W/cm^{2}}$ they exhibit a single peak at $k_{\Ion} \geq 0$ and not, as the EKSO solution, at $k_{\Ion} \leq 0$. In this intensity regime purely sequential double ionization dominates, pointing to possible shortcomings in the description of this process with both correlation potentials. 

As the general deficiencies of the uncorrelated functional described in the previous paragraph are entirely due to the functional for the observable, these results demonstrate the relative unimportance of the choice of the correlation potential in the TDKSE for the observables of interest in this work.

\subsection{Towards correlated functionals}
\label{Section Momentum Towards a Correlated Functional}
In polar representation, the solution of the TDSE is written as $\psi \ofxxt = \sqrt{\rho \ofxxt / \, 2} \ \euler^{\imagi \, \varphi \ofxxt}$ and the KS orbital as $\phi \ofxt = \sqrt{ n \ofxt / \, 2} \ \euler^{\imagi \, \vartheta \ofxt}$. We define a time-dependent complex exchange-correlation function
\begin{eqnarray}
\kappa \ofxxt & = & \frac{\psi \ofxxt}{\sqrt{2} \ \phi \ofxt[1] \, \phi \ofxt[2]} \nonumber \\
& = & \sqrt{ \gxc } \ \euler^{\imagi \, \left[ \varphi \ofxxt - \vartheta \ofxt[1] - \vartheta \ofxt[2] \right]} 
\label{Formula Momentum Kappa}
\end{eqnarray}
with the time-dependent exchange-correlation function $\gxc = \gxc \ofxxt$ given by $\gxc \ofxxt = \linebreak \rho \ofxxt / \, n \ofxt[1] \, n \ofxt[2]$. Approximations to $\gxc=\vert\kappa\vert^2$ have been used to construct correlated ionization probability functionals \cite{PetersilkaGross1999,WilkenBauer2006}. Note that while $\gxc$ is an observable (and thus a functional of only the density exists), the complex-valued $\kappa$ is not an observable. Using Eq.\,(\ref{Formula Momentum Kappa}) to express the correlated wavefunction $\psi \ofxxt$ in terms of the KS orbitals and the complex exchange-correlation function, Eq.\,(\ref{Formula Momentum Pair Density}) gives the correlated functional for the momentum pair density of the electrons freed in double ionization
\begin{eqnarray}
\lefteqn{ \rhotwoplus \ofkkt =  \pi^{-2} \, \left\vert \int \del x_{1} \int \del x_{2} \, \kappa \ofxxt \right.} & & \nonumber \\
& & \qquad \left. \times \, \phiplus \ofxt[1] \, \phiplus \ofxt[2] \, \euler^{- \imagi \, \left( k_{1} \, x_{1} + k_{2} \, x_{2} \right)} \right\vert^{2} 
\label{Formula Momentum Pair Density Correlated}
\end{eqnarray}
with $\phiplus \ofxt = f \ofx \, \phi \ofxt$. The correlated ion momentum density is calculated by using the correlated momentum pair density in Eq.\,(\ref{Formula Momentum Ion Density}). We thus have exact momentum distribution functionals, which depend only on the complex exchange-correlation function $\kappa$ and the KS orbital $\phi$. 

The complex exchange-correlation function $\kappa$ in turn depends on the pair density and the phase of the Schr\"odinger solution $\psi \ofxxt$. In order to derive momentum space properties for more complex atoms than Helium from the KS orbitals directly through expressions like Eq.\,(\ref{Formula Momentum Pair Density Correlated}), it is inevitable to approximate $\kappa$. However, this is challenging since, due to the Fourier-integrals in Eq.\,(\ref{Formula Momentum Pair Density Correlated}), the complex exchange-correlation function has to be approximated in all  $\mathcal{A} ( \Hetwoplus )$ (and not just for the bound electrons, as in the calculation of ionization probabilities \cite{PetersilkaGross1999,WilkenBauer2006}).

\subsection{Product phase approximation}
\label{Section Momentum Product Phase Approximation}
The necessary approximation of the complex exchange-correlation function $\kappa$ (\ref{Formula Momentum Kappa})  consists of approximating $\gxc \ofxxt$ and the phase-difference $\varphi \ofxxt - \linebreak \vartheta \ofxt[1] - \vartheta \ofxt[2]$.

Addressing the second part, the easiest approximation follows from the assumption that the difference of the sum of the phases of the KS orbitals and the phase of the correlated wavefunction can be neglected when calculating momentum distributions, i.e., we set 
\begin{equation}
\varphi \ofxxt = \vartheta \ofxt[1] + \vartheta \ofxt[2].
\label{Formula Momentum Phases PP}
\end{equation}
Since $\vartheta \ofxt$ is the phase of the KS orbital we denote this approach as the product phase (PP) approximation, which yields
\begin{equation}
\kappa^{\PP} \ofxxt = \sqrt{ \gxc \ofxxt } \, .
\label{Formula Momentum Kappa PP}
\end{equation}
It is noteworthy that knowledge of the exact $\kappa^{\PP}$ thus suffices to calculate the exact double ionization probabilities from the EKSO.

\begin{figure}[!tb]
\centering
\includegraphics[angle=0, width=0.45\textwidth]{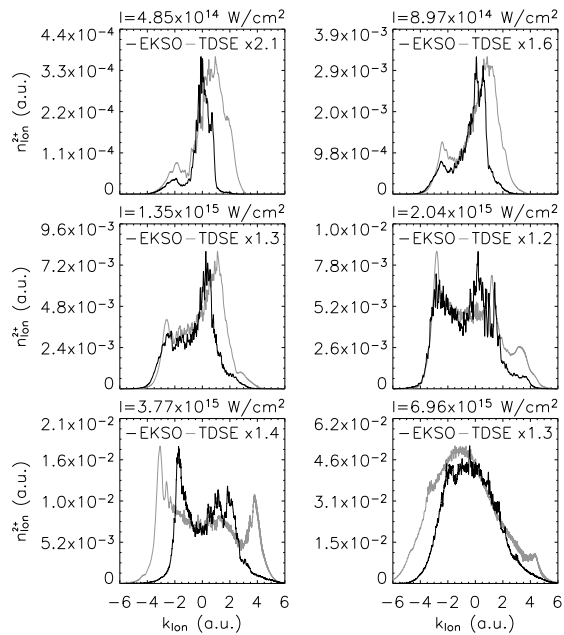}
\caption{Ion momentum density of the model $\Hetwoplus$ ion calculated from the correlated functionals in the PP approximation using the EKSO. Results for $\lambda \! = \! 780\,\nm$, $N \! = \!3$-cycle laser pulses with different effective peak intensities are compared to the ion momentum density obtained from the TDSE. \label{Figure Ion Momentum TDSE-EKSO-g}}
\end{figure}

We calculate the ion momentum density using Eq.\,(\ref{Formula Momentum Kappa PP}) in Eq.\,(\ref{Formula Momentum Pair Density Correlated}) and in Eq.\,(\ref{Formula Momentum Ion Density}). Employing the EKSO, the ion momentum densities shown in Fig.\,\ref{Figure Ion Momentum TDSE-EKSO-g} for $\lambda \! = \! 780\,\nm$, $N \! = \!3$-cycle laser pulses with different intensities are obtained. The results from the TDSE are depicted as well. For comparison of the qualitative features, they are scaled, although the integrals over the ion momentum densities are equal in both cases (note that the PP approximation returns the exact double ionization probabilities). A generally good qualitative agreement with the Schr\"odinger solution is acquired. The asymmetric structure and distinct peaks are reproduced. For intensities where NSDI is strongest, the quantitative agreement is least convincing. Although the PP approximation does not reproduce the exact $k_{\Ion}$ positions of the peaks, it modifies the uncorrelated functionals in a way which allows to deduce information about the underlying double ionization processes at the different intensities. We can therefore conclude that the difference between the phase of the correlated wavefunction and a product wavefunction is not as important for reproducing the structure of the ion momentum density as is the correlation given by $\gxc \ofxxt$. This conclusion was verified by setting $\gxc=1$ in Eq.\,(\ref{Formula Momentum Kappa}) and using the exact phases in Eq.\,(\ref{Formula Momentum Pair Density Correlated}), which did not yield the peaks present in the Schr\"odinger solution. Using LK05 orbitals in the PP approximation also reproduces distinct peaks while the general agreement with the Schr\"odinger ion momentum density is not as good as for the EKSOs.

The contour plots of the momentum pair density of the electrons freed in double ionization $\rhotwoplus \ofkkt$ calculated from the correlated functional in the PP approximation using the EKSO show a correlated structure, while differences from the TDSE momentum pair densities (Fig.\,\ref{Figure Momentum Pair Density TDSE-EKSO}) remain.

\begin{figure}[!tb]
\centering
\includegraphics[angle=0, width=0.45\textwidth]{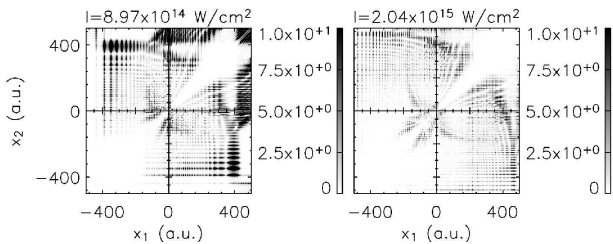}
\caption{Contour plots of the exchange-correlation function $\gxc \ofxxt$ for two effective peak intensities of $\lambda \! = \! 780\,\nm$, $N \! = \!3$ cycle laser pulses as acquired from the solution of the TDSE. For clarity values larger than $10$ are shown as $10$. \label{Figure Momentum gxc}}
\end{figure}

Using the PP approximation we obtain momentum distributions which yield fundamental insight into the double ionization processes. However, this still requires knowledge of the exact $\gxc \ofxxt$ at time $t=T$ after the laser pulse, i.e., of the exact pair density in real space. Approximating $\gxc \ofxxt$ is a formidable task itself. This can be seen from the highly correlated structure in Fig.\,\ref{Figure Momentum gxc} where contour plots of the exchange-correlation function $\gxc (x_{1},x_{2},T)$ are shown for intensities where NSDI dominates. An adiabatic approximation using the groundstate pair density \cite{WilkenBauer2006} is not feasible as the exchange-correlation function in the entire $\mathcal{A} ({\Hetwoplus}) $ is required in Eq.\,(\ref{Formula Momentum Pair Density Correlated}). An expansion for small inter-electron distances \cite{PetersilkaGross1999,Becke1988} will not include the correlations for large $\vert x_{1} - x_{2} \vert$, which are clearly present in Fig.\,\ref{Figure Momentum gxc}. By multiplying the complex exchange-correlation function with a damping function $F ( \vert x_{1} - x_{2} \vert )$ with $F \to 0$ for large $\vert x_{1} - x_{2} \vert$, we verified that short-range correlations alone in the final wavefunction do not reveal the characteristic peaks in the ion momentum density. It is therefore of central importance to devise new strategies of approximating $\gxc \ofxxt$. 

\section{Summary}
\label{Summary}
A model Helium atom in strong linearly polarized few-cycle laser pulses was investigated. Solution of the time-dependent Schr\"odinger equation yielded momentum pair distributions of the electrons freed in double ionization and corresponding ion momentum densities. They were consistent with a recollision process and, at higher laser intensities, with sequential double ionization. These results served as a reference for a  time-dependent density-functional treatment of the system. It was shown that the choice of the correlation potential in the Kohn-Sham equations is of minor importance compared to the form of the functionals for calculating the momentum distributions. An uncorrelated approach was found to produce ion momentum densities differing significantly from the Schr\"odinger solution in qualitative terms. We constructed an exact correlated functional via the two-electron wavefunction. The product-phase approximation reduces the problem of approximating this functional.

This work was supported by the Deutsche Forschungsgemeinschaft.


\begin{thebibliography}{23}
\expandafter\ifx\csname natexlab\endcsname\relax\def\natexlab#1{#1}\fi
\expandafter\ifx\csname bibnamefont\endcsname\relax
  \def\bibnamefont#1{#1}\fi
\expandafter\ifx\csname bibfnamefont\endcsname\relax
  \def\bibfnamefont#1{#1}\fi
\expandafter\ifx\csname citenamefont\endcsname\relax
  \def\citenamefont#1{#1}\fi
\expandafter\ifx\csname url\endcsname\relax
  \def\url#1{\texttt{#1}}\fi
\expandafter\ifx\csname urlprefix\endcsname\relax\def\urlprefix{URL }\fi
\providecommand{\bibinfo}[2]{#2}
\providecommand{\eprint}[2][]{\url{#2}}

\bibitem[{\citenamefont{Runge and Gross}(1984)}]{RungeGross1984}
\bibinfo{author}{\bibfnamefont{E.}~\bibnamefont{Runge}} \bibnamefont{and}
  \bibinfo{author}{\bibfnamefont{E.~K.~U.} \bibnamefont{Gross}},
  \bibinfo{journal}{Phys. Rev. Lett.} \textbf{\bibinfo{volume}{52}},
  \bibinfo{pages}{997} (\bibinfo{year}{1984}).

\bibitem[{\citenamefont{Marques et~al.}(2006)\citenamefont{Marques, Ullrich,
  Nogueira, Rubio, Burke, and Gross}}]{Marques2006}
\bibinfo{editor}{\bibfnamefont{M.~A.~L.} \bibnamefont{Marques}},
  \bibinfo{editor}{\bibfnamefont{C.~A.} \bibnamefont{Ullrich}},
  \bibinfo{editor}{\bibfnamefont{F.}~\bibnamefont{Nogueira}},
  \bibinfo{editor}{\bibfnamefont{A.}~\bibnamefont{Rubio}},
  \bibinfo{editor}{\bibfnamefont{K.}~\bibnamefont{Burke}}, \bibnamefont{and}
  \bibinfo{editor}{\bibfnamefont{E.~K.~U.} \bibnamefont{Gross}}, eds.,
  \emph{\bibinfo{title}{Time-Dependent Density Functional Theory}}
  (\bibinfo{publisher}{Springer}, \bibinfo{address}{Berlin Heidelberg},
  \bibinfo{year}{2006}).

\bibitem[{\citenamefont{Hohenberg and Kohn}(1964)}]{HohenbergKohn1964}
\bibinfo{author}{\bibfnamefont{P.}~\bibnamefont{Hohenberg}} \bibnamefont{and}
  \bibinfo{author}{\bibfnamefont{W.}~\bibnamefont{Kohn}},
  \bibinfo{journal}{Phys. Rev.} \textbf{\bibinfo{volume}{136}},
  \bibinfo{pages}{B864} (\bibinfo{year}{1964}).

\bibitem[{\citenamefont{Becker et~al.}(2005)\citenamefont{Becker, D\"{o}rner,
  and Moshammer}}]{BeckerDoerner2005}
\bibinfo{author}{\bibfnamefont{A.}~\bibnamefont{Becker}},
  \bibinfo{author}{\bibfnamefont{R.}~\bibnamefont{D\"{o}rner}},
  \bibnamefont{and}
  \bibinfo{author}{\bibfnamefont{R.}~\bibnamefont{Moshammer}},
  \bibinfo{journal}{J. Phys. B: At. Mol. Opt. Phys.}
  \textbf{\bibinfo{volume}{38}}, \bibinfo{pages}{S753} (\bibinfo{year}{2005}).

\bibitem[{\citenamefont{Fu et~al.}(2001)\citenamefont{Fu, Liu, Chen, and
  Chen}}]{FuLiu2001}
\bibinfo{author}{\bibfnamefont{L.-B.} \bibnamefont{Fu}},
  \bibinfo{author}{\bibfnamefont{J.}~\bibnamefont{Liu}},
  \bibinfo{author}{\bibfnamefont{J.}~\bibnamefont{Chen}}, \bibnamefont{and}
  \bibinfo{author}{\bibfnamefont{S.-G.} \bibnamefont{Chen}},
  \bibinfo{journal}{Phys. Rev. A} \textbf{\bibinfo{volume}{63}},
  \bibinfo{pages}{043416} (\bibinfo{year}{2001}).

\bibitem[{\citenamefont{Ho et~al.}(2005)\citenamefont{Ho, Panfili, Haan, and
  Eberly}}]{HoPanfili2005}
\bibinfo{author}{\bibfnamefont{P.~J.} \bibnamefont{Ho}},
  \bibinfo{author}{\bibfnamefont{R.}~\bibnamefont{Panfili}},
  \bibinfo{author}{\bibfnamefont{S.~L.} \bibnamefont{Haan}}, \bibnamefont{and}
  \bibinfo{author}{\bibfnamefont{J.~H.} \bibnamefont{Eberly}},
  \bibinfo{journal}{Phys. Rev. Lett.} \textbf{\bibinfo{volume}{94}},
  \bibinfo{pages}{093002} (\bibinfo{year}{2005}).

\bibitem[{\citenamefont{Lein and K\"ummel}(2005)}]{LeinKuemmel2005}
\bibinfo{author}{\bibfnamefont{M.}~\bibnamefont{Lein}} \bibnamefont{and}
  \bibinfo{author}{\bibfnamefont{S.}~\bibnamefont{K\"ummel}},
  \bibinfo{journal}{Phys. Rev. Lett.} \textbf{\bibinfo{volume}{94}},
  \bibinfo{pages}{143003} (\bibinfo{year}{2005}).

\bibitem[{\citenamefont{Wilken and Bauer}(2006)}]{WilkenBauer2006}
\bibinfo{author}{\bibfnamefont{F.}~\bibnamefont{Wilken}} \bibnamefont{and}
  \bibinfo{author}{\bibfnamefont{D.}~\bibnamefont{Bauer}},
  \bibinfo{journal}{Phys. Rev. Lett.} \textbf{\bibinfo{volume}{97}},
  \bibinfo{pages}{203001} (\bibinfo{year}{2006}).

\bibitem[{\citenamefont{V\'eniard et~al.}(2003)\citenamefont{V\'eniard,
  Ta\"{i}eb, and Maquet}}]{VeniardTaieb2003}
\bibinfo{author}{\bibfnamefont{V.}~\bibnamefont{V\'eniard}},
  \bibinfo{author}{\bibfnamefont{R.}~\bibnamefont{Ta\"{i}eb}},
  \bibnamefont{and} \bibinfo{author}{\bibfnamefont{A.}~\bibnamefont{Maquet}},
  \bibinfo{journal}{Laser Phys.} \textbf{\bibinfo{volume}{13}},
  \bibinfo{pages}{465} (\bibinfo{year}{2003}).

\bibitem[{\citenamefont{Ullrich et~al.}(2003)\citenamefont{Ullrich, Moshammer,
  Dorn, D\"{o}rner, Schmidt, and Schmidt-B\"{o}cking}}]{UllrichMoshammer2003}
\bibinfo{author}{\bibfnamefont{J.}~\bibnamefont{Ullrich}},
  \bibinfo{author}{\bibfnamefont{R.}~\bibnamefont{Moshammer}},
  \bibinfo{author}{\bibfnamefont{A.}~\bibnamefont{Dorn}},
  \bibinfo{author}{\bibfnamefont{R.}~\bibnamefont{D\"{o}rner}},
  \bibinfo{author}{\bibfnamefont{L.~P.~H.} \bibnamefont{Schmidt}},
  \bibnamefont{and}
  \bibinfo{author}{\bibfnamefont{H.}~\bibnamefont{Schmidt-B\"{o}cking}},
  \bibinfo{journal}{Rep. Prog. Phys.} \textbf{\bibinfo{volume}{66}},
  \bibinfo{pages}{1463} (\bibinfo{year}{2003}).

\bibitem[{\citenamefont{Bauer}(1997)}]{Bauer1997}
\bibinfo{author}{\bibfnamefont{D.}~\bibnamefont{Bauer}},
  \bibinfo{journal}{Phys. Rev. A} \textbf{\bibinfo{volume}{56}},
  \bibinfo{pages}{3028} (\bibinfo{year}{1997}).

\bibitem[{\citenamefont{Lappas and van Leeuwen}(1998)}]{LappasLeeuwen1998}
\bibinfo{author}{\bibfnamefont{D.~G.} \bibnamefont{Lappas}} \bibnamefont{and}
  \bibinfo{author}{\bibfnamefont{R.}~\bibnamefont{van Leeuwen}},
  \bibinfo{journal}{J. Phys. B: At. Mol. Opt. Phys.}
  \textbf{\bibinfo{volume}{31}}, \bibinfo{pages}{L249} (\bibinfo{year}{1998}).

\bibitem[{\citenamefont{Lein et~al.}(2000)\citenamefont{Lein, Gross, and
  Engel}}]{LeinGross2000}
\bibinfo{author}{\bibfnamefont{M.}~\bibnamefont{Lein}},
  \bibinfo{author}{\bibfnamefont{E.~K.~U.} \bibnamefont{Gross}},
  \bibnamefont{and} \bibinfo{author}{\bibfnamefont{V.}~\bibnamefont{Engel}},
  \bibinfo{journal}{Phys. Rev. Lett.} \textbf{\bibinfo{volume}{85}},
  \bibinfo{pages}{4707} (\bibinfo{year}{2000}).

\bibitem[{\citenamefont{Dahlen and van Leeuwen}(2001)}]{DahlenLeeuwen2001}
\bibinfo{author}{\bibfnamefont{N.~E.} \bibnamefont{Dahlen}} \bibnamefont{and}
  \bibinfo{author}{\bibfnamefont{R.}~\bibnamefont{van Leeuwen}},
  \bibinfo{journal}{Phys. Rev. A} \textbf{\bibinfo{volume}{64}},
  \bibinfo{pages}{023405} (\bibinfo{year}{2001}).

\bibitem[{\citenamefont{Bauer and Koval}(2006)}]{BauerKoval2006}
\bibinfo{author}{\bibfnamefont{D.}~\bibnamefont{Bauer}} \bibnamefont{and}
  \bibinfo{author}{\bibfnamefont{P.}~\bibnamefont{Koval}},
  \bibinfo{journal}{Comput. Phys. Comm.} \textbf{\bibinfo{volume}{174}},
  \bibinfo{pages}{396} (\bibinfo{year}{2006}).

\bibitem[{\citenamefont{Rudenko et~al.}(2004)\citenamefont{Rudenko, Zrost,
  Feuerstein, Jesus, Schr\"oter, Moshammer, and Ullrich}}]{RudenkoZrost2004}
\bibinfo{author}{\bibfnamefont{A.}~\bibnamefont{Rudenko}},
  \bibinfo{author}{\bibfnamefont{K.}~\bibnamefont{Zrost}},
  \bibinfo{author}{\bibfnamefont{B.}~\bibnamefont{Feuerstein}},
  \bibinfo{author}{\bibfnamefont{V.~L.~B.} \bibnamefont{Jesus}},
  \bibinfo{author}{\bibfnamefont{C.~D.} \bibnamefont{Schr\"oter}},
  \bibinfo{author}{\bibfnamefont{R.}~\bibnamefont{Moshammer}},
  \bibnamefont{and} \bibinfo{author}{\bibfnamefont{J.}~\bibnamefont{Ullrich}},
  \bibinfo{journal}{Phys. Rev. Lett.} \textbf{\bibinfo{volume}{93}},
  \bibinfo{pages}{253001} (\bibinfo{year}{2004}).

\bibitem[{\citenamefont{Liu et~al.}(2004)\citenamefont{Liu, Rottke, Eremina,
  Sandner, Goulielmakis, Keeffe, Lezius, Krausz, Lindner, Schatzel
  et~al.}}]{LiuRottke2004}
\bibinfo{author}{\bibfnamefont{X.}~\bibnamefont{Liu}},
  \bibinfo{author}{\bibfnamefont{H.}~\bibnamefont{Rottke}},
  \bibinfo{author}{\bibfnamefont{E.}~\bibnamefont{Eremina}},
  \bibinfo{author}{\bibfnamefont{W.}~\bibnamefont{Sandner}},
  \bibinfo{author}{\bibfnamefont{E.}~\bibnamefont{Goulielmakis}},
  \bibinfo{author}{\bibfnamefont{K.~O.} \bibnamefont{Keeffe}},
  \bibinfo{author}{\bibfnamefont{M.}~\bibnamefont{Lezius}},
  \bibinfo{author}{\bibfnamefont{F.}~\bibnamefont{Krausz}},
  \bibinfo{author}{\bibfnamefont{F.}~\bibnamefont{Lindner}},
  \bibinfo{author}{\bibfnamefont{M.~G.} \bibnamefont{Schatzel}},
  \bibnamefont{et~al.}, \bibinfo{journal}{Phys. Rev. Lett.}
  \textbf{\bibinfo{volume}{93}}, \bibinfo{pages}{263001}
  (\bibinfo{year}{2004}).

\bibitem[{\citenamefont{Rottke et~al.}(2006)\citenamefont{Rottke, Liu, Eremina,
  Sandner, Goulielmakis, Keeffe, Lezius, Krausz, Lindner, Sch\"atzel
  et~al.}}]{RottkeLiu2006}
\bibinfo{author}{\bibfnamefont{H.}~\bibnamefont{Rottke}},
  \bibinfo{author}{\bibfnamefont{X.}~\bibnamefont{Liu}},
  \bibinfo{author}{\bibfnamefont{E.}~\bibnamefont{Eremina}},
  \bibinfo{author}{\bibfnamefont{W.}~\bibnamefont{Sandner}},
  \bibinfo{author}{\bibfnamefont{E.}~\bibnamefont{Goulielmakis}},
  \bibinfo{author}{\bibfnamefont{K.~O.} \bibnamefont{Keeffe}},
  \bibinfo{author}{\bibfnamefont{M.}~\bibnamefont{Lezius}},
  \bibinfo{author}{\bibfnamefont{F.}~\bibnamefont{Krausz}},
  \bibinfo{author}{\bibfnamefont{F.}~\bibnamefont{Lindner}},
  \bibinfo{author}{\bibfnamefont{M.~G.} \bibnamefont{Sch\"atzel}},
  \bibnamefont{et~al.}, \bibinfo{journal}{J. of Mod. Opt.}
  \textbf{\bibinfo{volume}{53}}, \bibinfo{pages}{149} (\bibinfo{year}{2006}).

\bibitem[{\citenamefont{Figueira~de Morisson~Faria
  et~al.}(2004)\citenamefont{Figueira~de Morisson~Faria, Liu, Sanpera, and
  Lewenstein}}]{FariaLiu2004}
\bibinfo{author}{\bibfnamefont{C.}~\bibnamefont{Figueira~de Morisson~Faria}},
  \bibinfo{author}{\bibfnamefont{X.}~\bibnamefont{Liu}},
  \bibinfo{author}{\bibfnamefont{A.}~\bibnamefont{Sanpera}}, \bibnamefont{and}
  \bibinfo{author}{\bibfnamefont{M.}~\bibnamefont{Lewenstein}},
  \bibinfo{journal}{Phys. Rev. A} \textbf{\bibinfo{volume}{70}},
  \bibinfo{pages}{043406} (\bibinfo{year}{2004}).

\bibitem[{\citenamefont{Dreizler and Gross}(1999)}]{DreizlerGross1999}
\bibinfo{author}{\bibfnamefont{R.~M.} \bibnamefont{Dreizler}} \bibnamefont{and}
  \bibinfo{author}{\bibfnamefont{E.~K.~U.} \bibnamefont{Gross}},
  \emph{\bibinfo{title}{Density Functional Theory. An Approach to the Quantum
  Many-Body Problem}} (\bibinfo{publisher}{Springer}, \bibinfo{address}{Berlin
  Heidelberg}, \bibinfo{year}{1999}).

\bibitem[{\citenamefont{Pohl et~al.}(2000)\citenamefont{Pohl, Reinhard, and
  Suraud}}]{PohlReinhard2000}
\bibinfo{author}{\bibfnamefont{A.}~\bibnamefont{Pohl}},
  \bibinfo{author}{\bibfnamefont{P.-G.} \bibnamefont{Reinhard}},
  \bibnamefont{and} \bibinfo{author}{\bibfnamefont{E.}~\bibnamefont{Suraud}},
  \bibinfo{journal}{Phys. Rev. Lett.} \textbf{\bibinfo{volume}{84}},
  \bibinfo{pages}{5090} (\bibinfo{year}{2000}).

\bibitem[{\citenamefont{Petersilka and Gross}(1999)}]{PetersilkaGross1999}
\bibinfo{author}{\bibfnamefont{M.}~\bibnamefont{Petersilka}} \bibnamefont{and}
  \bibinfo{author}{\bibfnamefont{E.~K.~U.} \bibnamefont{Gross}},
  \bibinfo{journal}{Laser Phys.} \textbf{\bibinfo{volume}{9}},
  \bibinfo{pages}{105} (\bibinfo{year}{1999}).

\bibitem[{\citenamefont{Becke}(1988)}]{Becke1988}
\bibinfo{author}{\bibfnamefont{A.~D.} \bibnamefont{Becke}},
  \bibinfo{journal}{J. Chem. Phys.} \textbf{\bibinfo{volume}{88}},
  \bibinfo{pages}{1053} (\bibinfo{year}{1988}).

\end{thebibliography}
\end{document}